# Integrated and Spectrally Selective Thermal Emitters Enabled by Layered Metamaterials

Yongkang Gong[1], Kang Li [2,3] *, Nigel Copner[2,3], Heng Liu[4], Meng Zhao[4,] *, Bo Zhang[2, 5], Andreas Pusch[6], Diana L. Huffaker[1], and Sang Soon Oh[1,] *

[1]*School of Physics and Astronomy, Cardiff University, Cardiff, CF24 3AA, UK*

[2]*Wireless and Optoelectronics Research and Innovation Centre, Faculty of Computing, Engineering and Science, University of South Wales, Cardiff, CF37 1DL, UK*

[3]*Foshan Huikang Optoelectronics Ltd., B block, Sino-European Center, Foshan 528315, China*

[4]*Jiangsu Key Laboratory of Micro and Nano Heat Fluid Flow Technology and Energy Application, School of Physical Science and Technology, Suzhou University of Science and Technology, Suzhou, 215009, China*

[5]*Henan Academy of Special Optics Ltd., Xinxiang, 453000, China*

[6]*School of Photovoltaic and Renewable Engineering, UNSW, Sydney, New South Wales 2052, Australia*

**ABSTRACT**: Nanophotonic engineering of light-matter interaction at subwavelength scale allows thermal radiation that is fundamentally different from that of traditional thermal emitters and provides exciting opportunities for various thermal-photonic applications. We propose a new kind of integrated and electrically controlled thermal emitter that exploits layered metamaterials with lithography-free and dielectric/metallic nanolayers. We demonstrate both theoretically and experimentally that the proposed concept can create a strong photonic bandgap in the visible regime and allow small impedance mismatch at the infrared wavelengths, which gives rise to optical features of significantly enhanced emissivity at the broad infrared wavelengths of 1.4-14 µm as well as effectively suppressed emissivity in the visible region. The electrically driven metamaterial devices are optically and thermally stable at temperature up to ∼800 K with electro-optical conversion efficiency reaching ∼30%. We believe that the proposed high efficiency thermal emitters will pave the way towards integrated infrared light source platforms for various thermal-photonic applications and particularly provide a novel alternative for cost-effective, compact, low glare, and energy-efficient infrared heating.

## INTRODUCTION

Artificial control of thermal radiation that is difficult to attain with natural materials has been a research topic of interest for decades. The principle of manipulating thermal radiation is based on Kirchhoff's law, which states that the emissivity of an object is equal to its absorptivity for a given frequency, polarization, and direction. In the recent years, tremendous research efforts have been made toward tailoring light absorption based on plasmonic nanophononics attributed to the recent unprecedented development of nanofabrication techniques. Metals are usually known to be perfect reflectors but when they are structured on a scale of the wavelength, light reflection fades away and enhanced absorption occurs with a sharp spectrum much narrower than that of a blackbody due to excitation of resonant modes confined in the subwavelength metallic cavities or excitation of surface plasmon polaritons (SPPs) on the corrugated metal surfaces. Various types of spectrally selective narrowband nanophotonic absorbers have been developed such as nanogratings,[1-4] photonic crystals,[5-7] and three-layer-metamaterials.[8-11] These narrowband absorbers/emitters have triggered promising applications in many different areas ranging from optical sensors,[12-14] hot electron photodetectors,[15] optical modulators,[16-20] high speed switching,[21] energy recycling,[22-24] image encryption,[25, 26] to thermal imaging.[27]

In addition to the narrowband absorbers, recently there has been a strong motivation to enhance the light–matter interaction with a broadband absorption response by artificially manipulating the effective permittivity and permeability of nanophotonic structures to control the resonant modes. A number of strong broadband absorber schemes have been investigated, for example, by exploiting refractory metasurfaces,[28-33] metal-insulator-metal nanostructures,[34-36] semiconductor photonic crystals,[37, 38] and multilayer thin films.[39-44] The broadband absorbers/emitters attract increased attention in fundamental science and have found a number of excited applications such as solar energy,[45-47] thermophotovoltaics,[31, 48, 49] infrared stealth,[50] and radiative cooling.[40, 41, 51]

In this paper, we propose and demonstrate a new kind of electrically controlled layered-metamaterial thermal emitters (LTEs) composed of multiple dielectric/metallic nanolayers. In contrast to the other reported nanophotonic thermal emitters, the proposed LTEs are thermal-photonic integrated and have advantage of tailoring thermal radiation with selectively enhanced emissivity in a broadband infrared regime and effectively suppressed emissivity at the visible wavelengths. We design and analyze the optical characteristics of the LTEs both analytically and numerically, and experimentally investigate their optical and thermal properties including angular dependent emissivity, spectral radiation, thermal photonic properties, and electro-optical conversion efficiency. Our study offers a cost-effective, spectrally selective, and integrated infrared light source strategy that could find various applications. For example, it provides a solution that could possibly overcome the drawbacks of the existing infrared heater technologies (i.e., dazzling glare and low emissivity at the infrared wavelengths).[52]

**RESULTS**

*Structure and design metrology.* The proposed LTEs have a configuration featuring two stacked one-dimensional (1D) periodic lattices, as is illustrated in [Figures 1](a-c), where a finite periodic lattice (named as (Si/Cr/Si)$^n$) with unit cell consisting of triple nanolayers of Si, Cr and Si deposited on top of a $Ni_{80}Cr_{20}$ thin film on a quartz substrate. On top of this lattice is another finite periodic lattice (denoted as $(SiO_2/Si)^m$) with two alternately arranged nanolayers of $SiO_2$ and Si. Here, $m$ and $n$ represent the number of the periods of the two lattices. The reason we choose $SiO_2$ and Si materials for the top lattice is because of their high refractive index difference and high melting point. The former property enables broad photonic bandgap generation, while the latter allows the proposed LTEs to operate at the desired elevated temperature. The $SiO_2$ layer is on top of the Si layer in the unit cell of the lattice $(SiO_2/Si)^m$, which makes our structure with a cap layer $SiO_2$ to protect the whole structure from oxidation. Electrical voltage is applied to the $Ni_{80}Cr_{20}$ thin film to generate Joule heat to raise the temperature of the LTEs. We particularly choose $Ni_{80}Cr_{20}$ as the electrically driven metallic layer, because $Ni_{80}Cr_{20}$ material has high electrical resistivity and is efficient for generating Joule heat. The $Ni_{80}Cr_{20}$ layer also acts as a light reflector and allows little transmitted light. Its thickness does not affect the optical properties of the LTEs in the considered wavelength range as long as it is thicker than hundreds of nanometers.

We model and design the proposed LTEs by starting with optimizing the top photonic lattice $(SiO_2/Si)^m$ to make the LTEs with low emissivity at the visible wavelengths. We investigate the optical spectra of the structure by the transfer matrix method (TMM) with the refractive indices of all materials taken from experimental data,[53] and optimize the optical spectra by adjusting the nanolayers' thickness and the number of the periods of the lattice $(SiO_2/Si)^m$ $m$ (for detailed design and optimization, see Section A in Supporting information). When the thickness of thin film $SiO_2$ (Si) is 100 nm (40 nm) and $m$ is 4, broadband and high reflectivity at the wavelengths of 0.45-0.75 $\mu$m is achieved in a broad angle of incidence (AOI) ranging from 0 degree up to 80 degrees (see the inset of Figure 1d and Figure S1 in Supporting Information). According to the International Commission on Illumination (CIE),[54] there are three infrared radiation bands (i.e., IR-A 0.7-1.4 $\mu$m, IR-B 1.4-3 $\mu$m, and IR-C 3-1000 $\mu$m) and since the IR-B and IR-C bands are particularly important for many real-world applications such as chemical/medical sensing and infrared heating, we focus on designing the LTEs with enhanced emissivity at wavelengths of 1.4-14 $\mu$m. To this end, we fix the geometric parameters of the optimized lattice $(SiO_2/Si)^m$ aforementioned, and then optimize the lattice $(Si/Cr/Si)^n$ to minimize the characteristic impedance difference $|Z_{\text{LTEs}} - Z_{\text{air}}|$ to reduce reflectivity in the infrared regime and at the same time retain high reflectivity of the LTEs at the visible wavelengths, by adjusting the films' thicknesses and the number of lattice periods $n$ (for the detailed optimization, see Equation S1 and Figure S2 in Supporting Information). Here, $Z_{\text{air}}$ and $Z_{\text{LTEs}}$ are the characteristic impedance of air and the LTEs, respectively. We observe from Figure 1d that the impedance mismatch at the visible wavelengths is larger than that of the infrared wavelengths when the thickness of Si (Cr) film is 100 nm (4 nm) and $n$ is 6, indicating the light reflection at visible wavelengths dominates the infrared light reflection. This is verified by the reflectivity spectra numerically calculated by TMM, as depicted in Figures 1e, where high reflectivity at the wavelengths of 0.45-0.75 $\mu$m and low reflectivity at the wavelengths of 1.4-14 $\mu$m is achieved. This selectively ultrabroadband reflectivity characteristic enables LTEs with high absorptivity in the infrared regime and low absorptivity in the visible regime, considering the transmissivity of the proposed LTEs is tiny at all the considered wavelengths (Figure 1f).

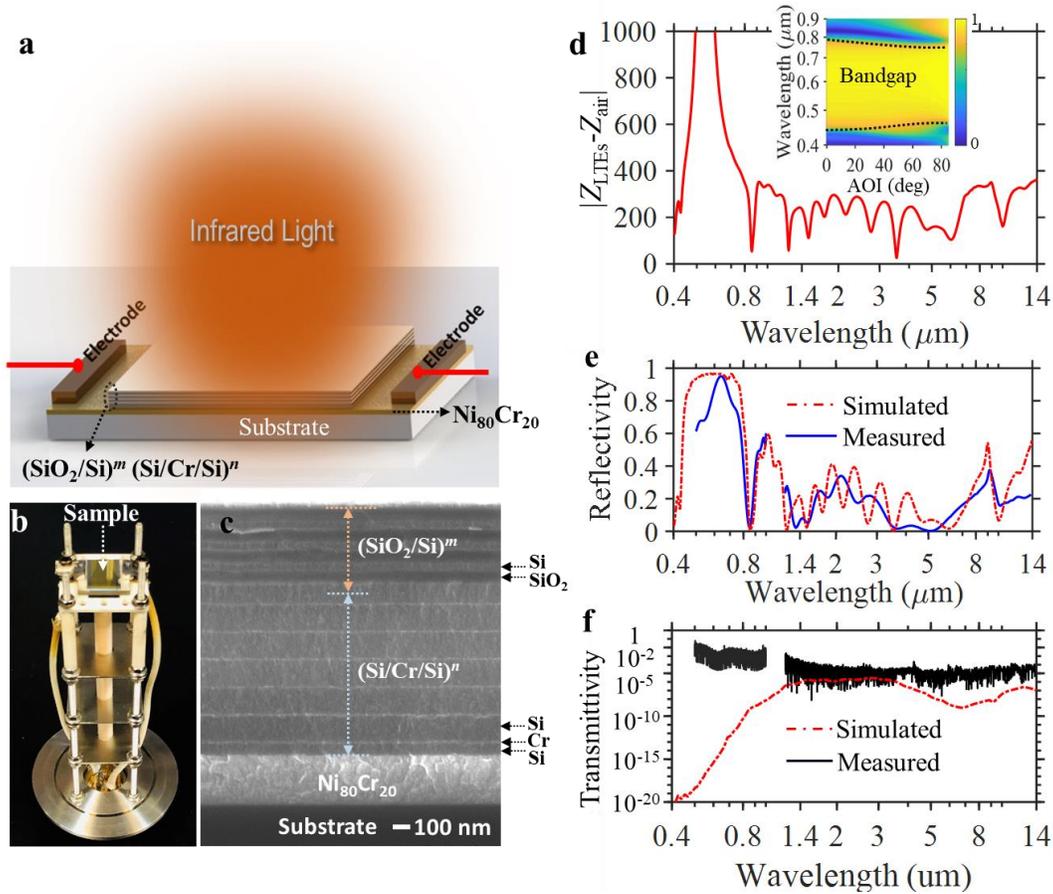

**Figure 1**. The concept, and the theoretical and experimental implementation of the selectively broadband metamaterial thermal emitters. (a) Schematic diagram of the LTEs, where two 1D photonic lattices with structures of $(Si/Cr/Si)^n$ and $(SiO_2/Si)^m$ are lying on top of a $Ni_{80}Cr_{20}$ nanolayer deposited on a quartz substrate. Voltage is applied to the $Ni_{80}Cr_{20}$ layer to generate Joule heat to raise the temperature of the whole device. Here, $m$ and $n$ represent the number of the periods of the two lattices. (b) Photography image of the fabricated devices mounted on a sample holder. (c) SEM image of the cross section of the LTEs. (d-f) The characteristic impedance mismatch $|Z_{LTEs}-Z_{air}|$, and the reflectivity and the transmissivity spectra, respectively. The inset of (d) depicts the reflectivity spectra versus angle of incidence (AOI) of the $(SiO_2/Si)^m$. The geometric parameters of the LTEs are: the thickness of each Si ($SiO_2$) layer in the $(SiO_2/Si)^m$ is 40 nm (100 nm), and the thickness of each Si (Cr) layer in the $(Si/Cr/Si)^n$ is 100 nm (4 nm); the number of periods $m$ and $n$ are 4 and 6, and the thickness of the $Ni_{80}Cr_{20}$ layer and the substrate is 300 nm and 0.5 mm, respectively.

*Device fabrication and optical spectra measurements.* We fabricate the designed LTEs using E-beam evaporation and measure angle-dependent reflectivity and transmissivity spectra by Fourier-transform infrared spectroscopy (FTIR) and grating spectrometer (for fabrication and measurement details, see Sections B and C in Supporting Information). The measured reflectivity and transmissivity spectra are consistent with the theoretical prediction, as demonstrated in Figures 1e and 1f. Based on the reflectivity and transmissivity, we obtain the spectral hemispherical absorptivity $A(\theta, \lambda)$ of the LTEs for both transverse-electric (TE) and transverse-electric (TM) polarized light, where $\theta$ and $\lambda$ is AOI and wavelength, respectively. The absorptivity has no azimuthal angle dependence and only has polar angle dependence due to the one-dimensional geometry of the LTEs. It is noted from Figures 2a-b that that both TE and TM polarized light experience high (low) absorptivity in a broad spectral and angular range at the infrared (visible) wavelengths. The angle-dependent emissivity of the LTEs is derived by averaging the TE- and TM-polarized absorptivity. The simulated and measured emissivity spectra at various angles from 5° to 75° are plotted and compared in Figure 2c, showing a good agreement between our theory and experiment. Figure 2c indicates that in comparison with the typical thermal emitter metals (such as Ni, Cr and W) that suffer from high (low) emissivity in the visible (infrared) regime, the proposed LTEs have advantage of suppressed emissivity at visible wavelengths and enhanced emissivity at the infrared wavelengths with a broad spectral and angular response. At the angle of $\theta = 5°$, for example, the calculated emissivity is as high as ~0.81 averaged over wavelengths of 1.4-14 $\mu$m, and is as low as ~0.07 averaged over wavelengths of 0.45-0.75 $\mu$m. Emissivity peaks at wavelength of ~10 $\mu$m is noticed at all the angles, as indicated by the vertical lines in Figure 2c. This is due to high light absorption induced by the large imaginary part of the refractive index of the SiO$_2$ thin films at this wavelength.

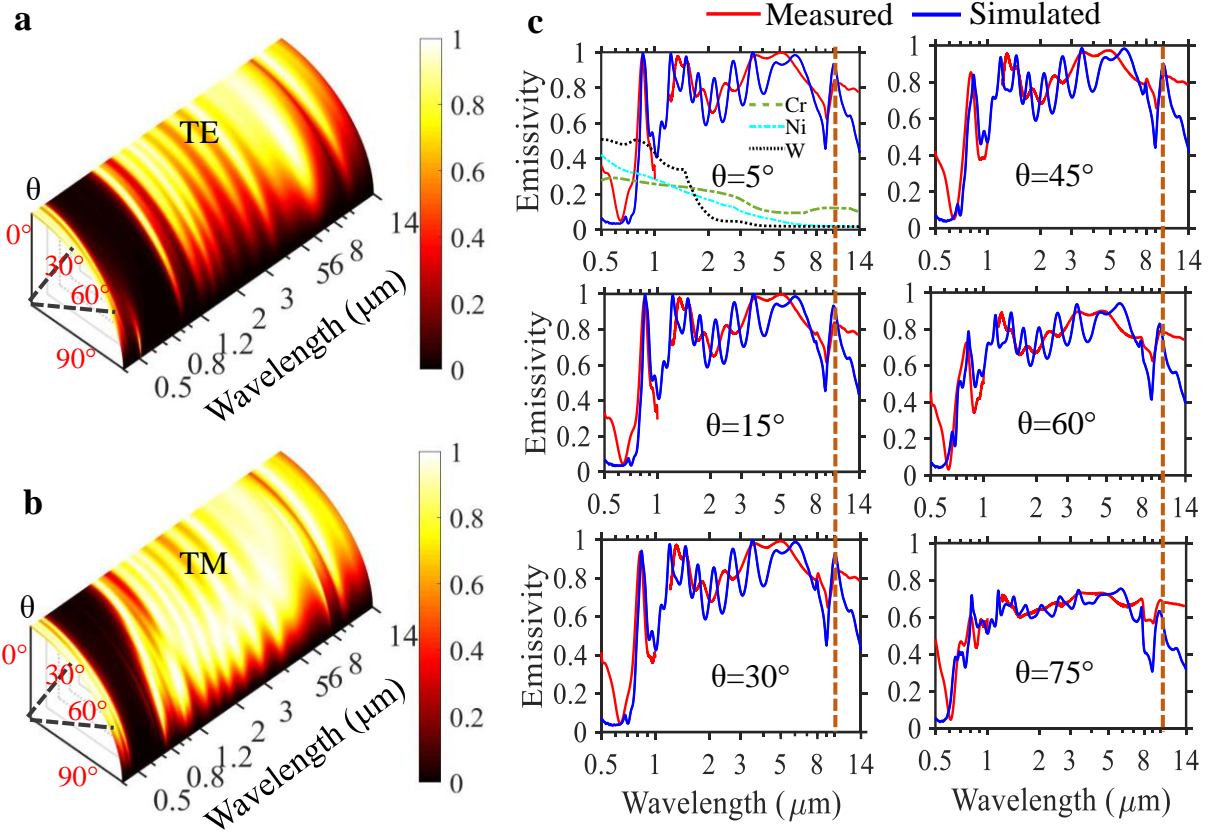

**Figure 2**. The spectral characteristics of the LTEs at different angle and polarization. (a), (b) The calculated absorptivity spectra versus the angle of incidence for the TE- and TM-polarized light, respectively. (c) The measured and calculated emissivity spectra at various angles from 5° to 75°. The calculated emissivity is obtained by averaging the TE- and TM-polarized absorptivity spectra from (a), while the measured emissivity is derived from collecting variable angle specular reflectance with un-polarized light illumination. The emissivity spectra of the LTEs are compared with that of a 300 nm-thick thin films of refractory metals (tungsten (W), nickel (Ni) and chromium (Cr)) that are widely used for thermal emitters, demonstrating that the proposed structures offer enhanced (suppressed) emissivity in the infrared (visible) regime. The vertical lines in (c) indicate the presence of an emissivity peak at ~10 $\mu$m. The structure geometric parameters are the same as in Figure 1.

***Emissivity and thermal radiation at high temperature***. An important question is whether the thermal emitters are optically and thermally stable at high temperatures. To this end, we fabricate LTEs with different structure dimension of width *w* and length *l* (see Figure 3a) and undertake a series of high temperature measurements. We measure the spectral emissivity and thermal radiation of the LTEs at different operating temperatures controlled by the electric current flowing through the Ni$_{80}$Cr$_{20}$ thin film (for detailed experimental measurements, see Section C in Supporting Information). The generated Joule heat depends on both the input electrical voltage and the structure resistance that relies on the size of the Ni$_{80}$Cr$_{20}$ film. We measure the I-V curve of the fabricated LTEs with different structure size and show that the structure resistance decreases with *w* and increases with *l* (see Figure S3 in Supporting Information). The thermal image of the electrically heated LTEs taken by an infrared camera (Figure 3b) shows uniform temperature distribution on the surface with slightly decreased temperature at the edges due to thermal convection from the structure to the ambient environment. We evaluate the optical stability of the LTEs by measuring the emissivity spectra at different operating temperatures. Figure 3c clearly demonstrates that the emissivity does not degrade much at high temperatures especially at the infrared wavelengths, although a slight increase of the emissivity is observed at the visible wavelengths due to the expected increase of the electron collision frequency at high temperatures, which leads to increased free carrier absorption.

To investigate the thermal radiation property, we collimate and direct the radiated light of LTEs into a FTIR to measure the radiated power as function of the operating temperature of the LTEs (for the detailed experimental setup, see Sections C and D in Supporting Information). Figure 3d illustrates that the thermal radiation becomes stronger when the device temperature increases, which is consistent with simulated results in Figure 3e. A spectral peak at ~10 $\mu$m is observed in both measurements and simulations, as marked by the vertical dashed line in Figures 3d-e, which is due to the high emissivity peak around this wavelength (see the vertical dashed lines in Figure 2c). The discrepancies between the measurements and simulations are caused by several factors. For example, the surface temperature of the LTEs is not perfectly uniform as assumed in simulations, and it is impossible to well collimate the radiated light at all the wavelengths to the detector for measurements since the LTEs radiate light covering broad wavelengths. We also compare the thermal radiation of LTEs with an ideal blackbody and show that the proposed LTEs give rise to radiation spectra that are very close to that of the blackbody at the same operating temperature (Figure 3e).

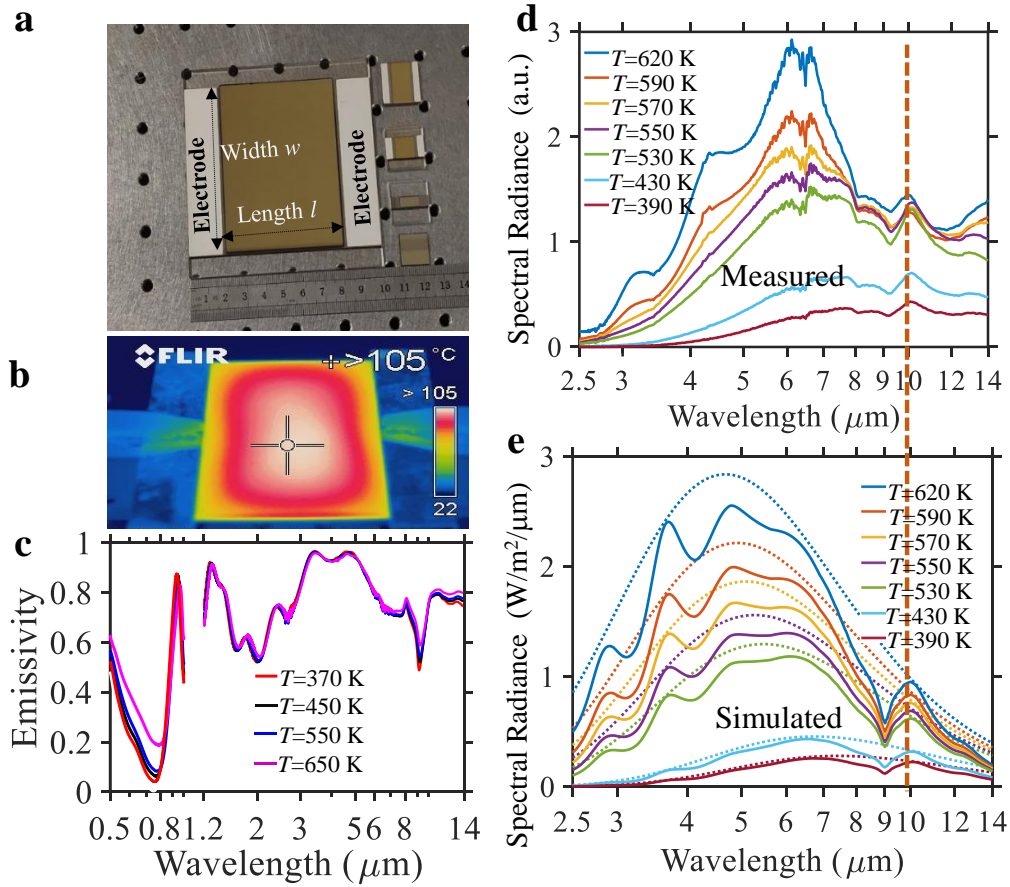

**Figure 3**. The optical emissivity and thermal radiation of the LTEs at the elevated temperature. (a) A photograph of the fabricated LTEs with different width $w$ and length $l$. (b) The thermal distribution of the samples taken by an infrared camera. (c) The measured emissivity at the angle of $\theta = 45°$ under different structure temperature obtained by controlling the voltage applied to the $Ni_{80}Cr_{20}$ layer. (d) The measured spectral radiation intensity versus the device temperature by directing the radiated light into a parabolic mirror to be collimated into a FTIR for detection. (e) The calculated spectra radiance (solid lines) compared to that of an idea blackbody (dotted lines) at the same operating temperature. The calculations are performed by integrating the emitted light within divergence angle of $5°$ from the normal of the LTEs (See Equation S2 in Supporting Information). Both the measured and calculated spectra show a peak at $\sim 10$ $\mu$m (marked by the vertical dashed line) due to strong emissivity at this wavelength (see the vertical dashed line in Figure 2c). The structure geometries are the same as those in Figure 2.

***Properties of the thermal photonic dynamics.*** To fully characterize the thermal photonic performance of LTEs, it is essential to explore how it responds to the input electricity in real-time. To do this, we gradually increase the voltage applied to the $Ni_{80}Cr_{20}$ layer with ramp rate of 0.1 V per 15 mins (see the inset of Figure 4a) to ensure there is enough time for the temperature to increase and get stabilized. At the same time, we monitor the evolution of the current, the surface temperature, and the radiated power. We place the electrically controlled LTEs in a vacuum chamber and put it very close to the KBr window of the chamber (for the detailed measurement system, see Section E and Figure S4 in Supporting Information). The radiation from the heated LTEs passes through the KBr window and enters the input port of a gold-coated integrating sphere that sits very closely to the other side of the KBr window, and a photodetector is placed at the exit port of the integrating sphere to collect the light and measure the power. We see from Figure 4a that the current, the device surface temperature and the radiated power (measured at the exit port of the integrating sphere) increases with the applied voltage until the temperature reaches ~800 K. When the voltage is further increased, the nanolayers of the LTEs crack and the LTEs are not conductive any longer. As a result, the current, the temperature, and the radiated power drop dramatically. We calibrate the integrating sphere with a commercial blackbody and measure the blackbody radiation power $P_1$ at the input port of the integrating sphere and the light power $P_2$ at the exit port of the integrating sphere, and derive the transmission efficiency of the integrating sphere by $\eta = P_2/P_1 = $ ~0.065 % (see Figure S6 in Supporting Information). By normalizing the light power in Figure 4a to $\eta$, we get the power $P_o$ of the light radiated from the LTEs. The evolution of $P_o$ and the surface temperature to the input electricity power $P_e$ is given in Figure 4b. Based on the results, we obtain the electro-optical conversion efficiency of the LTEs by extracting $P_o$ versus $P_e$. Figure 4c shows that the conversion efficiency reaches ~30% at temperature > 700 K. The conversion efficiency can be further increased by minimizing the heat dissipation, for example, by reducing the thickness of the device substrate.

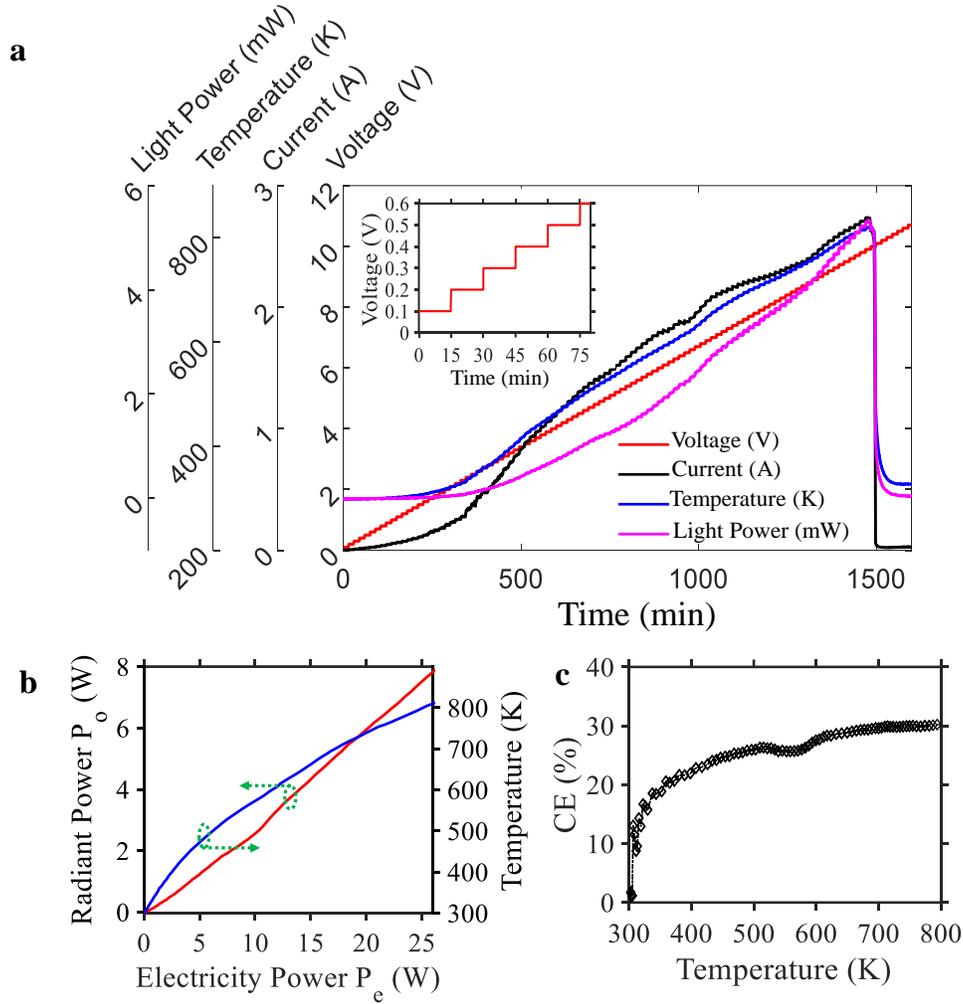

**Figure 4**. The thermal photonic dynamics of the electrically controlled LTEs. (a) The evolution of the current, the surface temperature and the radiated power (detected at the exit port of the integrating sphere) to the input electrical voltage that increases with a step of 0.1 V per 15 minutes. The inset plots the zoom-in of the voltage versus time. The light radiated from the LTEs propagates through the KBr window of the vacuum chamber and enters the integrating sphere through the input port, and then is detected by a photodetector at the exit port of the integrating sphere. (b) The dependence of the radiant light power $P_o$ (at the input port of the integrating sphere) on the input electrical power $P_e$ and the surface temperature of the LTEs. (c) The electro-optical conversion efficiency $CE = P_o/P_e$ of the LTEs. The thickness of the structure nanolayers are the same as those in Figure 3, and the structure width is $w = 20$ mm and length is $l = 10$ mm.

## CONCLUSIONS

We have proposed an integrated thermophotonic layered-metamaterial scheme to obtain efficient and spectrally selective thermal emitters with greatly enhanced emissivity in the infrared regime and effectively suppressed emissivity in the visible regime. With our optimized thermal emitters, we have achieved an averaged emissivity as high as ~0.81 over an broad infrared wavelength range of 1.4-14 $\mu$m and an averaged emissivity as low as ~0.07 at visible wavelengths of 0.45-0.75 $\mu$m. We have experimentally verified the proposed concept by exploring the optical and thermal characteristics including the angular and temperature-dependent emissivity, the spectral radiation, the thermal dynamics, and the electro-optical conversion efficiency. Our measurement results demonstrate that proposed LTE devices are optically and thermally stable up to ~800 K and yield an electro-optical conversion of ∼30%. Another advantage of the LTEs is the feature of one-dimensional structural simplicity, which allows large-scale production with low cost fabrication. We believe the LTEs offer an enhanced and integrated infrared source strategy that could find promising applications such as infrared heating, thermophotovoltaics, and thermal imaging etc. Finally, we stress that the thermal photonic performance of the LTEs could be further improved of by incorporating the proposed concept with other dielectric and metal materials (such as tungsten, hafnium dioxide, and aluminum oxide) that have better thermal and mechanical stability.

## AUTHOR INFORMATION

**Corresponding Authors**

* Email: kang.li@southwales.ac.uk

* Email: mzhao@usts.edu.cn

* Email: OhS2@cardiff.ac.uk

**Author Contributions**

Y.G. conceived the project. Y.G., K.L., and N.C. performed designs and numerical simulations. K.L. and B.Z. implemented device fabrications. Y.G. and K.L. set up the system for the emissivity and reflectivity spectra measurement. M.Z. and H.L. built the high-temperature characterization setup and undertook characterization of the device thermal photonic radiation. Y.G. wrote the manuscript with major contributions from S.S.O. and A.P. All authors discussed the simulated and measured results and the manuscript. All authors have given approval to submission of the manuscript.

## ACKNOWLEDGMENTS

The authors gratefully acknowledge Zhibo Li and Shiyu Xie for their assistance in the device characterization and valuable discussions. We thank Rui Dong and Dominic Kwan for the help with the FTIR measurements. This work is supported by 2015 Foshan Technology Innovation Group project (Advanced Solid-State Light Source Application and Innovation Team) and European Regional Development Fund through the Welsh Government (80762-CU145 (East)).

52. Space heating is a huge source of energy consumption. In comparison with the air space heating technologies (i.e., heat is generated by electricity/gas and dissipates into air) that is very inefficient since heat convects to the whole space, infrared heating provides optical heating locally and instantly and has reported to be more energy efficient. The existing commercial infrared panels (typically made of refractory metals such as tungsten, nichrome alloys, and ceramic materials etc), however, have two major issues that have inhibited the market penetration to industry and homeowners: dazzling glare and low efficiency. The glare occurs due to strong radiation in visible wavelengths, which not only causes light pollution. The low efficiency arises from low optical emissivity of the infrared panels at the infrared wavelengths. Enhancing infrared heating performance relies on material innovation of increasing emissivity in infrared regime and decreasing emissivity at visible wavelengths. .

53. Palik, E. D., *Handbook of optical constants of solids*, Academic press, **1998**.

54. International Commission on Non-Ionizing Radiation Protection, ICNIRP guidelines on limits of exposure to incoherent visible and infrared radiation. *Health Phys.* **2013**, *105,* 74-96.

Supporting Information

# Integrated and Spectrally Selective Thermal Emitters Enabled by Layered Metamaterials

Yongkang Gong, Kang Li, Nigel Copner, Heng Liu, Meng Zhao, Bo Zhang, Andreas Pusch, Diana L. Huffaker, and Sang Soon Oh

## A. Design strategy and device optimization

Our layered-metamaterial thermal emitter concept allows integrated thermophotonic devices that can significantly enhance emissivity in a broadband infrared regime and at the same time effectively suppress emissivity in the visible regime. According to Kirchhoff's law, the emissivity is determined by the absorptivity, hence we can tailor the absorptivity of the LTEs to get the desired thermal radiation properties. The proposed concept utilizes two 1D finite photonic lattices of $(SiO_2/Si)^m$ and $(Si/Cr/Si)^n$ as well as a $Ni_{80}Cr_{20}$ thin film to effectively optimize the characteristic impedance mismatch $|Z_{\text{LTEs}} - Z_{\text{air}}|$ to reduce light reflectivity and increase absorptivity at the wavelengths of interested, as depicted in Figure 1 in the main text. The characteristic impedance of the LTEs is given by recursive relation of [1]

$$\xi_{k+1} = Z_{k+1} \frac{\xi_k - i Z_{k+1} \tanh(\beta_{k+1} d_{k+1})}{Z_{k+1} - i \xi_k \tanh(\beta_{k+1} d_{k+1})}, \tag{S1}$$

where the subscript $k$ ($k = 1, 2 \dots, 2m + 3n + 1$) represents the $k^{\text{th}}$ layer of the LTEs counted from the top layer of the LTEs. $\beta_k$ is the wave vector and $d_k$ and the thickness of the $k^{\text{th}}$ layer. $Z_{\text{air}} = \xi_0$ and $Z_{\text{LTEs}} = \xi_{2m+3n+1}$ are the characteristic impedance of air and the whole LTE structure, respectively. Here, $m$ and $n$ represent the number of periods of the two lattices.

Firstly, we design and optimize the $(SiO_2/Si)^m$ at the visible wavelengths. We particularly choose $SiO_2$ and Si materials for the grating because of their large refractive index difference, which allows generation of a broad photonic bandgap. We calculate the optical spectra with the transfer matrix method (TMM)[2] using measured refractive indices for all the materials. As shown in Figure S1a, a broadband photonic bandgap appears and covers wavelengths of 0.45-0.75 $\mu$m when the thickness of the Si ($SiO_2$) thin films is 40 nm (100 nm). It is noted that the reflectivity increases with period $m$ until $m$ reaches 4. When $m$ is larger than 4, the reflectivity hardly increases, and the photonic bandgap becomes narrower. Therefore, we choose $m = 4$ as the optimized parameter. Figure S1b indicates that the $(SiO_2/Si)^m$ has tiny absorptivity in the infrared regime except at wavelengths of 9-10 $\mu$m due to the large imaginary part of refractive index of $SiO_2$ material in this wavelength range.

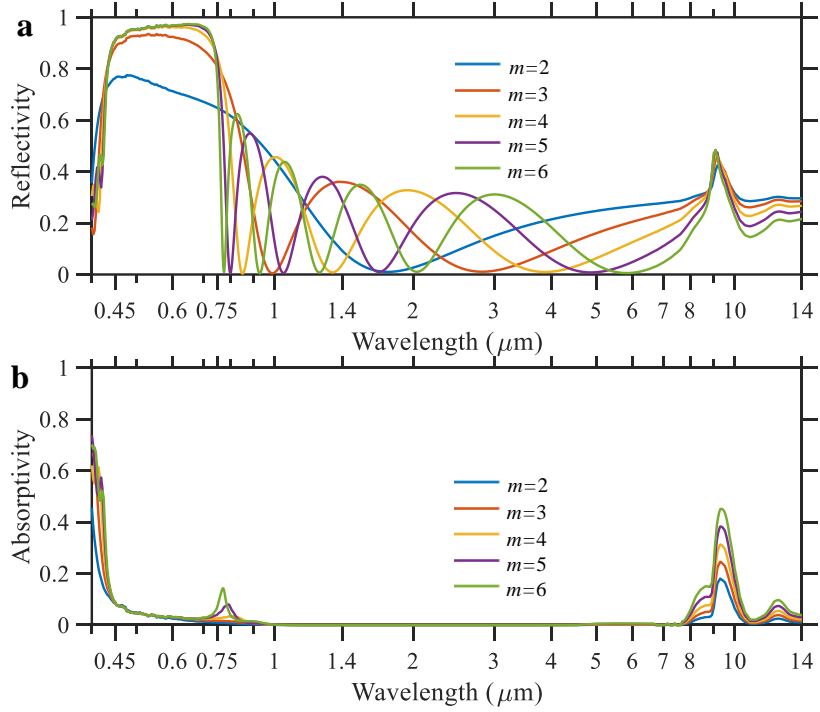

**Figure S1**. The optimization of the finite photonic lattice $(SiO_2/Si)^m$ to have strong and broadband reflectivity in the visible wavelength regime. (a) Reflectivity and (b) absorptivity spectra of the lattice $(SiO_2/Si)^m$ versus the number of periods $m$. The lattice exhibits low absorptivity in the infrared regime, showing that the enhanced infrared emissivity characteristic of the proposed LTEs arises from interaction among the lattice $(SiO_2/Si)^m$, the lattice $(Si/Cr/Si)^n$ and the nanolayer $Ni_{80}Cr_{20}$. For the optimized structure, the thickness of $SiO_2$ and Si in the unit cell of lattice $(SiO_2/Si)^m$ is 100 nm and 40 nm, respectively, and the number of periods $m$ is 4. The optimization of the lattice $(Si/Cr/Si)^n$ is depicted in Figure S2.

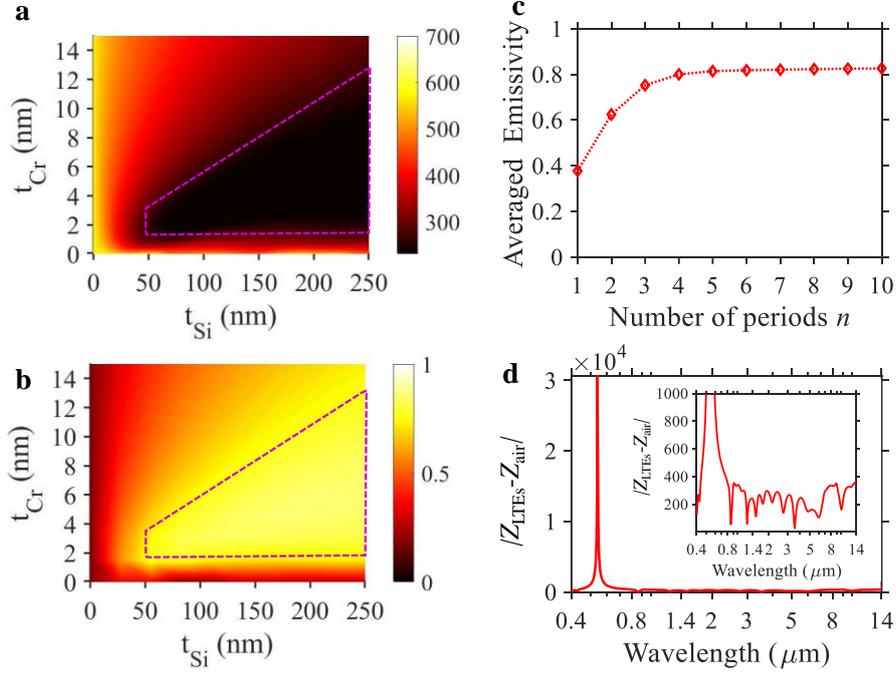

**Figure S2.** The optimization of the finite photonic lattice $(Si/Cr/Si)^n$ to have strong emissivity (absorptivity) of the proposed LTEs in the infrared regime. (a), (b) Dependence of the averaged impedance mismatch $|Z_{LTEs} - Z_{air}|$ and the averaged emissivity of the LTEs on film thickness $t_{Si}$ and $t_{Cr}$, respectively. Here, the impedance mismatch and emissivity are averaged over the wavelength range of 1.4-14 $\mu$m. $t_{Si}$ and $t_{Cr}$ are the thickness of Si and Cr in the unit cell of the $(Si/Cr/Si)^n$. The impedance mismatch analysis in (a) agrees well with the emissivity spectra obtained by TMM in (b), indicating that strong infrared emissivity occurs at the film thickness within the region marked by dashed lines. (c) The averaged emissivity of the LTEs versus the number of the lattice periods $n$. (d) $|Z_{LTEs} - Z_{air}|$ versus wavelengths when $t_{Si} = 100$ nm, $t_{Cr} = 4$ nm, and $n = 6$. The other film thickness in above simulations is the same as those in Figure S1.

Based on the optimized lattice $(SiO_2/Si)^m$, we then optimize the lattice $(Si/Cr/Si)^n$ to minimize the impedance mismatch $|Z_{LTEs} - Z_{air}|$ to achieve enhanced emissivity of the LTEs in the infrared regime. To this end, we calculate the averaged $|Z_{LTEs} - Z_{air}|$ over the wavelength range of 1.4-4 $\mu$m at various $t_{Si}$ and $t_{Cr}$ based on Equation S1, where $t_{Si}$ and $t_{Cr}$ are the

thickness of the Si and Cr nanolayers in each unit cell of the $(Si/Cr/Si)^n$, respectively. We observe from Figure S2a that the smallest impedance mismatch occurs at the structure parameters within the marked area, which is validated by the absorptivity spectra of the LTEs obtained by TMM as shown in Figure S2b. It shows that high absorptivity can be achieved when the impedance mismatch is small. The averaged emissivity versus the number of periods $n$ is plotted in Figure S2c. It is observed that the average emissivity increases with $n$ and reaches an almost stable value of ~0.81 when $n$ is larger than 5. Based on above analysis, we choose the optimized parameters of $t_{Si} = 100$ nm, $t_{Cr} = 4$ nm, and $n = 6$. We observe from Figure S2d that these parameters enable LTEs with large impedance mismatch in visible wavelengths and small impedance mismatch in the infrared range of 1.4-14 μm, which gives rise to selective broadband reflectivity characteristic (see Figure 1e in the main text).

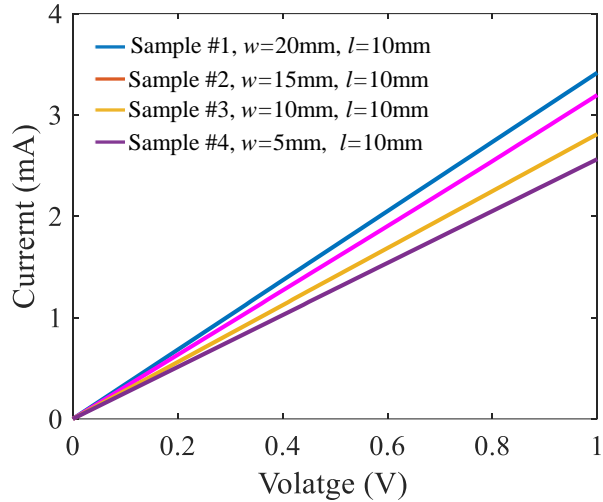

**Figure S3**. The measured I-V curve of the fabricated LTEs with different width $w$ and length $l$ (see Figure 3a). The structure parameters are: the thickness of Si (SiO$_2$) layer in the unit cell of the $(SiO_2/Si)^m$ is 40 nm (100 nm), and the thickness of Si (Cr) layer in the unit cell of the $(Si/Cr/Si)^n$ is 100 nm (4 nm). The number of lattice periods $m$ and $n$ are 4 and 6, respectively, and the thickness of the Ni$_{80}$Cr$_{20}$ film is 300 nm.

**B. Fabrication of the designed LTEs**

We fabricate the designed LTE devices by electron-beam evaporation of the constituent materials onto polished planar quartz substrates with 0.5 mm thickness. The thickness of each nanolayer during the material evaporation is measured with an in-situ quartz crystal monitor. Imaging of the nanolayer thickness is also performed by ultrahigh-resolution SEM (Hitachi Regulus 8220) scanning of the cross sections of the fabricated samples (Figure 1c). Conductive metal paste is applied to the edges of the NiCr layer as electrodes (see Figure 1a and Figure 3a). LTE devices with different dimensions are fabricated (Figure 3a and Figure S3). The I-V characteristics of the devices are measured using a probe station (Cascade Mps 150) and are plotted in Figure S3, illustrating that changing structure dimension can vary the electrical resistance and hence control the Joule heat generated from the LTEs.

## C. Angular dependent reflection spectra measurements

We use a fiber-coupled light source (Thorlabs SLS202/M) to illuminate the fabricated devices that are placed on a rotary stage, and utilize a spectrometer (Ocean Optics USB-4000) to collect the angle-dependent reflected light spectra at the wavelength range of 0.5-1 μm. A Fourier transform infrared (FTIR) spectrometer (Nicolet iS50) with variable angle specular reflectance accessory (Harrick Seagull) is used to measure reflection spectra at wavelengths of 1.2-14 μm with angles of incidence varying from 5 deg up to 75 deg. In the two types of measurements, the light reflection of the samples is normalized against a broadband silver mirror reference to obtain reflectivity spectra. Since the transmission of the LTE is negligible (Figure 1f), we can derive the absorptivity and emissivity (Figure 2) directly from the measured reflectivity. For the emissivity spectral measurements at high temperature (Figure 3c), electric current from a DC power supplier passes through the NiCr layer of the LTEs to generate Joule heat to raise the temperature of the whole structure. A sensitive thermocouple and infrared camera detect the device surface temperature (Figure 3b).

## D. Spectral thermal radiation

For the spectral thermal radiation measurements at various temperatures (Figure 3d), the radiation from the electrically heated LTEs is collimated by a 90° off-axis parabolic mirror with protected gold coating (Thorlabs MPD249-M01) toward the external port of the FTIR and is collected by a DTGS-KBr detector. To compare the measured results with theory, we calculate the spectral radiance $E(\lambda, T)$ by integrating the radiation within the angle of radiation $\theta_0$ from the normal, i.e.,

$$E(\lambda, T) = 2\pi \int_0^{\theta_0} \varepsilon(\theta, \lambda) \frac{2hc^2}{\lambda^5 (e^{\frac{hc}{\lambda kT}} - 1)} \sin(\theta) \cos(\theta) d\theta \tag{S2}$$

where $h = 6.626 \times 10^{-34}$ J·S, $k = 1.38 \times 10^{-23}$ J·K$^{-1}$, and $c$ are Planck constant, Boltzmann constant, and speed of light in vacuum, respectively. $T$ is the device temperature, and $\varepsilon(\theta, \lambda)$ is the device emissivity at radiation angle $\theta$ and wavelength $\lambda$. We calculate the spectral radiance based on Equation S2 and plot it in Figure 3e. Clearly, the measured spectral power results are consistent with the calculated results, demonstrating that the radiated light power increases with the operating temperature of the LTEs. The reason for the discrepancy between the measured and calculated results are explained in the main text.

### E. Characterization of thermophonic dynamics, the total radiated power and electro-optical conversion efficiency

A vacuum-based automatic measurement system is built to undertake the thermophonic dynamic measurements (Figure S4). Figure S5 shows that the vacuum system can effectively reduce heat convection loss of the LTE devices. In the system, automation modules of a programmable DC power supply and a control unit are controlled by LabVIEW software to simultaneously record the applied voltage, the chamber pressure, the current flowing through samples, and the operating temperature and the radiation power of the samples. The DC power supply feeds the samples with gradually increased voltage with a ramp rate 0.1 V per 15 mins (Figure 4a). The LTE samples are mounted on the sample holder (Figure 1b) and are inserted into the vacuum chamber that is pumped by a dry scroll pump. The structure and materials of the sample holder are carefully designed to minimize thermal conduction loss. Two gold-plated electrodes are used to apply the DC power to the samples, and a thermocouple is used to measure the surface temperature of the samples via a point contact. The electrically heated samples are placed very closely to the chamber's KBr window that is cooled by a circulating chiller to prevent over-heating. The radiated light from the heated samples passes though the KBr window and enters an a customized two-ports gold-coated integrating sphere that sits very close to the other side of the KBr window. The KBr window has a transmittance >90% in the wavelength range of 0.4-20 µm. The integrating sphere has a high diffuse reflectance of >98% in the wavelength range of 0.7-20 µm. A sensitive thermopile-based photodetector (spectral range 0.19-20 µm, Ophir Photonics) with a power meter is placed at the exit port of the integrating sphere to detect the radiated power.

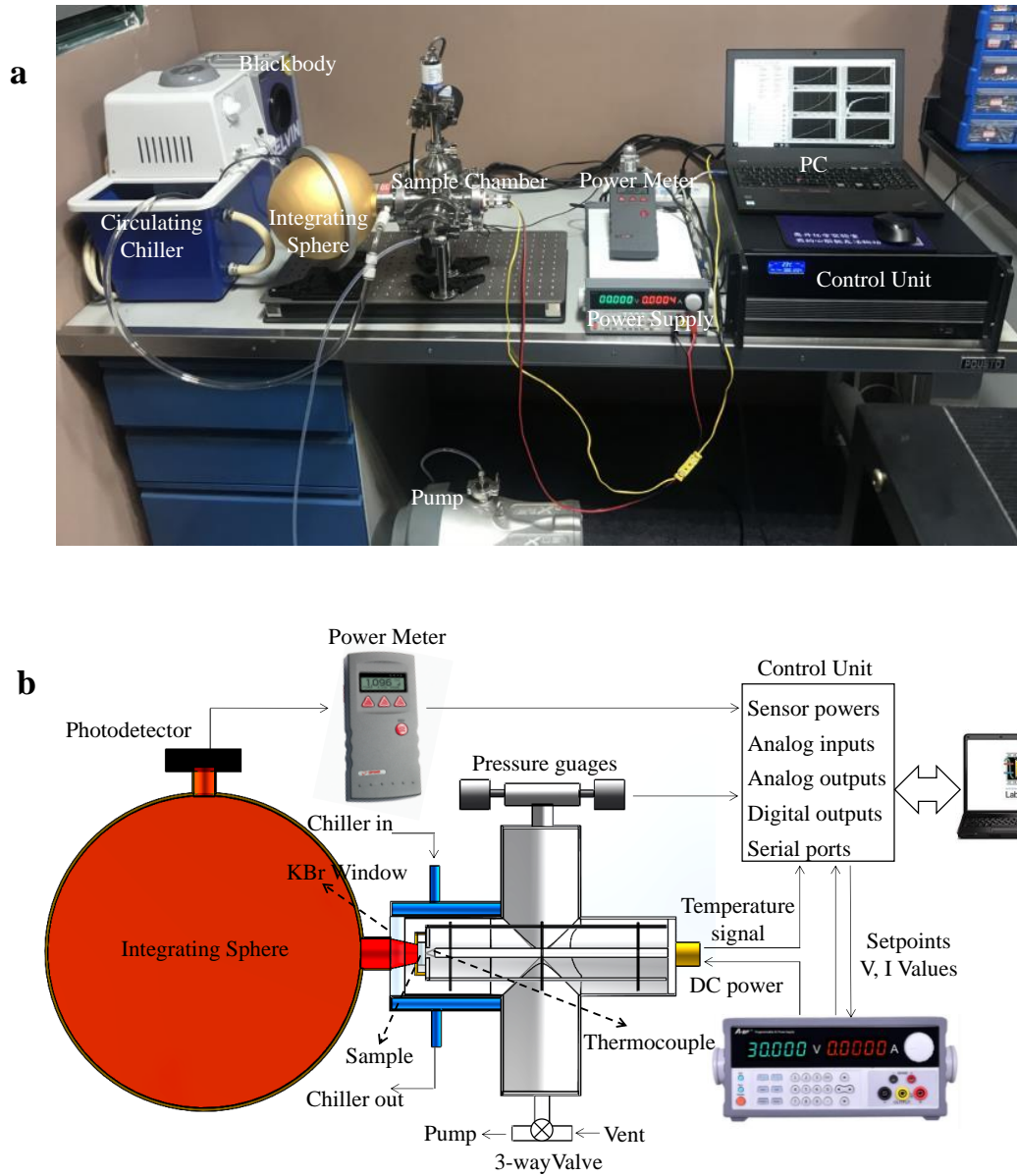

**Figure S4**. (a) Photography image of the developed thermophonic measurement system to characterize the fabricated LTE devices. (b) The schematic diagram of the measurement system. The LTE samples in this measurement have width of 20 mm and length of 10 mm.

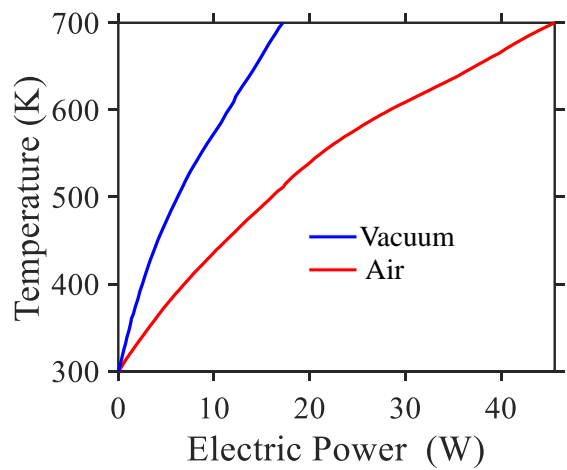

**Figure S5**. The measured surface temperature of the LTEs placed in vacuum and air, respectively. The device temperature is higher in vacuum than that in air (i.e., the ambient room environment) under the same input electric power due to the reduced heat convection loss. The LTE device in this measurement has width of 20 mm and length of 10 mm. The thickness of the nanolayers is the same as those in Figure S3.

To ensure the accuracy of the optical power measurement, we use a commercial blackbody as the light source to calibrate the light transmission efficiency from the input to the exit ports of the integrating sphere. Based on the measured radiated light power $P_1$ and $P_2$ of the blackbody at the input and the exit ports, we obtain the transmission efficiency of the integrating sphere by $\eta = P_2/P_1 = \sim 0.065$ %, as shown in Figure S6. The total radiated power of the LTEs can be estimated by normalizing the power of radiated light collected at the exit port of the integrating sphere to the light transmission efficiency of the integrating sphere η (see Figures 4a-b in the main text).

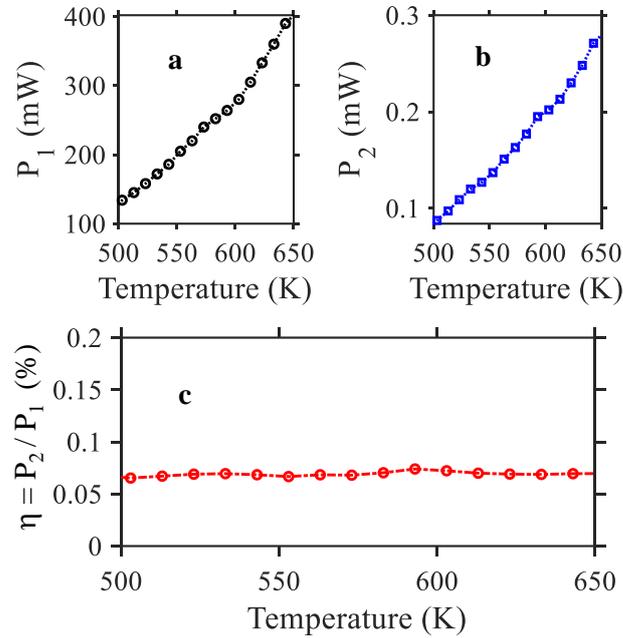

**Figure S6.** Calibration of the light transmission efficiency from the input to the exit ports of the integrating sphere with a commercial blackbody as the light source. The radiated light power $P_1$ measured at the input (a) and the power $P_2$ measured at the exit port (b) of the integrating sphere versus the operating temperature of the blackbody. (c) The light transmission efficiency of the integrating sphere obtained by $\eta = P_2/P_1$.